\documentclass{svjour3}
\smartqed  
\usepackage{natbib}

\usepackage{graphicx}

\journalname{Scientometrics}
\title{On the Internal Dynamics of the Shanghai Ranking}
\author{Domingo Docampo \and Lawrence Cram}
\institute{Domingo Docampo \at
              Universidad de Vigo,   Atlantic Research Center for Information and Communication Technologies; Campus Universitario, 36310 Vigo, Spain.\\
              Tel.: +34-986-812134 \email{ddocampo@uvigo.es}
\and
Lawrence Cram \at
              Australian National University,   University House; 1 Balmain Crescent, Acton ACT 0200, Australia.\\
              Tel.: +61-2-6125-5334 \email{Lawrence.Cram@anu.edu.au}
              }
              \date{Received: date / Accepted: date}
\begin{document}


\maketitle

\begin{abstract}
The Academic Ranking of World Universities (ARWU) published by researchers at Shanghai Jiao Tong University has become a major source of information for university administrators, country officials, students and the public at large. Recent discoveries regarding its internal dynamics allow the inversion of published ARWU indicator scores to reconstruct raw scores for five hundred world class universities. This paper explores raw scores in the ARWU and in other contests to contrast the dynamics of rank-driven and score-driven tables, and to explain why the ARWU ranking is a score-driven procedure. We show that the ARWU indicators constitute sub-scales of a single factor accounting for research performance, and provide an account of the system of gains and non-linearities used by ARWU. The paper discusses the non-linearities selected by ARWU, concluding that they are designed to represent the regressive character of indicators measuring research performance.  We propose that the utility and usability of the ARWU could be greatly improved by replacing the unwanted dynamical effects of the annual re-scaling based on raw scores of the best performers.

\keywords{University\and Ranking \and ARWU\and Dynamics  \and Shanghai \and Score}

\end{abstract}
\section{Introduction}
 \label{intro}

International rankings constitute benchmarking tools that enable performance comparisons among academic institutions. Rankings attract the interest of the public at large, and higher education officials understand that academic institutions tend to be compelled by rankings to be more accountable, set strategic planning goals, and provide comparative information to students, parents and other stakeholders \citep{Hazelkorn}. 	

Almost from its inception in 2003, the Shanghai Jiao Tong University Institute of Higher Education academic ranking of world universities (ARWU) has stimulated high levels of interest, use, debate, controversy and emulation. ARWU ranks the research performance of academic institutions on the basis of numerical measures of research quality and quantity \citep{Butler2010}. The  indicators relate either to research production or to excellence recognized by prestigious awards or by a high number of citations, and are open to public scrutiny. Over the past ten years the ranking ``... has attracted a great deal of attention from the scientific community worldwide, in part due to the simplicity and transparency of its criteria'' \citep{zitt}.

The importance of the Shanghai ranking has been recognized by governments and university administrators. An influential European education think tank, Bruegel, acknowledged the impact of ARWU in the foreword of one of its policy papers on Higher Education in Europe, ``$\ldots$ the `Shanghai ranking' has set in motion a major re-examination of higher education policies throughout Europe. It has also triggered reform initiatives aimed at fostering excellence and recognition, illustrating again the potency of benchmarking" \citep{aghion2008}. The results from the Shanghai ranking have been used to assess the research strengths and shortcomings of national higher education systems, normalizing by population (\citeauthor{aghion2008}, \citeyear{aghion2008}; \citeauthor{aghion2010}, \citeyear{aghion2010}), by share of world's GDP \citep{docampo2011}, and by share of world's economic capacity \citep{marginson2007}.

The Shanghai ranking has become a major resource for exploring  the characteristics and quality of academic institutions and university systems worldwide. The accessibility of the sources of the raw data and the fact that the hierarchy of universities generated by ARWU roughly aligns with perceptions of the historical and recent performance of elite research universities have contributed to its acceptance and success.

University ranking on an international scale is a relatively recent phenomenon, and methodological problems are not unexpected \citep{sawyer}. However, discussions around ARWU and other ranking systems tend to be either so highly critical that any potential usefulness is dismissed, or so willingly accepting that obvious flaws are overlooked. Faced with this polarisation alongside the evident acceptability of ARWU, we concur with the opinion of the late Alan Gilbert, onetime vice-chancellor of the University of Manchester: ``all current university rankings are flawed to some extent; most, fundamentally, but rankings are here to stay, and it is therefore worth the time and effort to get them right'' - quoted in  \cite{Butler2007}. 
From this perspective there is a need for accounts of the Shanghai ranking that attend to pragmatic aspects.

By the term pragmatic, we imply a focus on the way that the current ARWU raw data are processed, in contrast to the task of generating or re-generating new categories of raw data. The so-called Leiden ranking \citep{waltman} rises to the challenge of generating new categories of data. It relies wholly on bibliometric information, removing the Nobel Prize and Fields Medal indicators of the ARWU. The bibliometric components of the Leiden ranking also differ in many details from the ARWU ranking including (1) replacement of the highly-cited researcher category with counts of an institution�s highly cited publications, (2) adoption of fractional counts for collaborative publications, and (3) the open possibility of exploring non-English publications. The price of the Leiden methodology compared with the ARWU methodology is its reliance on powerful analytical methods built up by a large team over many years. It is difficult enough for researchers to explore the ARWU methodology: reproducing and extending the detail of the Leiden methods would seem to be beyond the reach of most other researchers in the field. There are evident benefits of ``open'' ranking systems that are not in effect locked to a single provider, and for this reason we have adopted a pragmatic view of the ARWU raw data in this study.

One common criticism of ARWU is that it combines disparate indicators into a single total score, and that it does this wrongly. \cite{billaut} for example assert that the ``aggregation technique used is flawed". However, the problem of combining disparate indicators of performance or status into a single list is not confined to global rankings of universities. Well-established methods for disparate aggregation that are widely accepted in other fields include consumer preference voting systems and combined event sporting tables. Issues in such systems include the mapping of performance to score within each event or category, the correspondence of performance and scores across the different events or categories, and the balancing of the components in the aggregate \citep{grammaticos}. The Shanghai ranking addresses the first point by providing a precise procedure to assign scores to performance on the six indicators (sub-scales). The six sub-scales are clearly disparate (e.g. there are many nulls in the Alumni and Award indicators) but this feature is a deliberate inclusion \citep{liucheng}. While the selection of the indicator weights in the ARWU cannot be uniquely benchmarked since there are no agreed measures of research quality, there is nevertheless a straightforward connection of the ARWU scale with the quality of the research carried out in an academic institution \citep{ioannidis}. The measurement validity, tantamount to the reliability of the raw scores on the indicators, is no longer questionable since the Shanghai ranking results can be accurately reproduced \citep{docampo2013}.

This paper is a comparative study of the internal dynamics of the Shanghai ranking and scoring system, and similar systems used in other contexts. By {\it dynamics} we mean the transformation of the raw performance data and its combination into the final list of scores and rankings. Significant differences in nomenclature exist across different ranking systems. There is some risk that our use of terms such as `contestants' and `events' could be misunderstood as a viewpoint that the universities ranked in ARWU have somehow agreed to enter a competition they strive to win. There are profound differences between the socialisation processes that have established the measures of heroic human physical striving that characterise peak athletics performance, and the {\em post facto} construction of ostensible measures of human intellectual striving that in practice trivializes the valuation of scholarly activity.

The outline of the paper is as follows. First we summarise the raw data of the ARWU and its transforming dynamics. We then outline a taxonomy of rankings and league tables, drawing attention to the differences between rank-driven and score-driven approaches. ARWU dynamics fit within the score-driven category, and additionally can be classified among the three basic types of score-driven tables: linear, progressive and regressive. Whereas combined events in track and field use a progressive point table \citep{trkal}, the Shanghai ranking is a regressive point table owing to the convergence of raw scores among the lower-ranked universities. We conclude by analysing the dependence of the ARWU ranking scoring system on the best performer, and suggesting that the current ARWU approach of annually adjusted scaling should be replaced by a new approach using fixed gains.  This would remove an unnecessary and undesirable dependence of scores on the performance of other institutions, and readily support monitoring of the evolution of the indicator scores for all universities including many not currently included in the ARWU ranking.

 \section{\label{methodology}ARWU ranking}

ARWU ranks universities individually or into bands by sorting on the total score. The total score is the linearly weighted sum of $6$ indicator scores derived from the corresponding raw data by transformations that have been calibrated and explained by \cite{docampo2013}.  A first hand account of the Shanghai ranking methodology can be found in \cite{liucheng}. For the purposes of this paper it is useful to consider a {\em procedural} description of the steps required to produce the annual ARWU report as follows:

\begin{enumerate}

\item {\it Assemble new raw data}

Scan the relevant data sources \citep{liucheng} for updated information on the {\it alma mater} and employing institution of new Nobel prize winners and Fields medalists, new awardees and revised affiliations for Thompson ISI Highly Cited researchers, and the relevant institutional details of Nature \& Science publications and ISI indexed publications in the prior year. Resolve as far as practicable issues relating to the legal definitions of the institutions that are being tracked as potential members of the top $500$ and decide how these definitions relate to the many institutional aliases that will have been used as pointers to these new data. Scan official data sets when published to obtain the relevant institutional FTE (full-time equivalent) faculty numbers used in the calculation of the PCP (per-capita performance) indicator.

\item {\it Assess and determine exceptions}

Once per decade, rescale indicators for historical Nobel prize winners and Fields medalists according to the decadal aging procedure. Allocate institutions specialized in the humanities and the social sciences to the class that uses special weights to exclude Nature \& Science publications. Decide which entire countries and which special institutions will be excluded from the determination of PCP using published FTE.

\item {\it Combine raw data and apply scaling}

Sum and/or average raw data over the relevant time windows according to the published algorithms. For every indicator other than PCP, multiply each value of the processed raw data by a fixed scaling factor so that Harvard University has a scaled raw score of $10000$. Calculate an intermediate quantity for PCP by dividing the weighted sum of the scaled raw scores by the FTE for the California Institute of Technology, and apply a scaling factor so that this intermediate quantity for CalTech is $10000$.

\item {\it Compress the scaled raw data}

For each indicator, including PCP, compress the dynamic range of the scaled raw data by taking its square root to form the indicator score.

\item {\it Calculate the total score and determine ranking/banding}

Combine the indicator scores using the relevant indicator weights, and linearly re-scale so that the total score for Harvard University is $100$. Rank institutions by total score, and determine for publication the ranking sequence, the membership of bands, and the excluded institutions.

\end{enumerate}

\noindent We return to discuss the dynamics of these procedures following a review of ranking taxonomies.

\section{Ranking Taxonomies}
\setcounter{figure}{0}

In generic terms, rankings and league tables that sequence an overall performance measure of separate contestants in different events can be classified into two major types, rank-driven or score-driven, depending on whether they combine the ranking positions or actual marks across events. Both methods find practical application in a wide variety of settings, and both have strengths and weaknesses that tend to favour one or the other in a particular setting.

\subsection{Rank-driven sequences}

Rank-driven sequencing aggregates the ranked positions of the different contestants over the various event indicators to form a single overall performance indicator. A simple example is the calculation of the team score in a cross-country race as the sum of the finishing positions of a fixed number of team members. If the event indicators do not induce reliable integer sequences (for example, because of subjective dimensions of event marking, or measurement error, or truncated marks in some events) non-integer event indicators may be sequenced, clustered or banded into ordered sequences. This procedure is used by the World Bank and similar organisations in a number of socio-economic assessments. \cite{dasgupta} use the approach to measure the quality of life in nations and \cite{worldbank} to measure the ease of doing business. Ranking and clustering algorithms used in rank-driven sequencing are designed to harmonise disparate measures and distributions of marks before aggregation by converting them to quasi-uniform distributions that retain no information about the original measured values beyond their sequence. Although the measures are originally quantitative and disparate, a rank-driven sequence becomes {\it agnostic} about the actual measured values \citep{sawyer}.

Rank-driven sequencing relates to the Borda rule and its variants \citep{myerson}. These are methods of rank-order voting or sequencing widely used in the political and social sciences in settings such as the resolution of polls or the reporting of aggregated public opinion over a range of topics. Borda rule rankings suffer from a number of drawbacks, including the peculiar feature known as Arrow's paradox - in which the ranking of two contestants (A ahead of B) can be reversed merely by the inclusion of an apparently irrelevant third contestant (\citeauthor {Arrow}, \citeyear{Arrow}; \citeauthor{Saari}, \citeyear{Saari}; \citeauthor{Hammond}, \citeyear{Hammond}). Dependence on irrelevant results can be avoided by alternative ways of resolving the overall sequence, as in the `count gold first' rule used to generate the rank-driven sequencing of national medal counts in Olympic games.

Variants of the Borda Rule could be used in the ARWU. One important choice would be the treatment of the abundant zero counts for Nobel prize winners and Fields medalists: choosing between assigning zero count or the score prescribed by the rule. To illustrate, let us indicate what would happen if the World Bank Knowledge Assessment Methodology (KAM) \citep{worldbank} were used in ARWU: First, $500$ universities would be ranked in order from `best' to `worst' on each indicator using their actual scores. Their scores would then be normalized on a scale of $0 $ to $10$ following the KAM procedure:
\begin{enumerate}
\item For each indicator $i$, and for a particular university $n$, identify the number of institutions with higher rank $(N_n)$.
\item Compute the score $I_n$ on the indicator following the normalization formula: $I_n = 10(1-N_n/500)$
\end{enumerate}

\begin{figure}[htbp]
\includegraphics[height=16cm,scale=0.5]{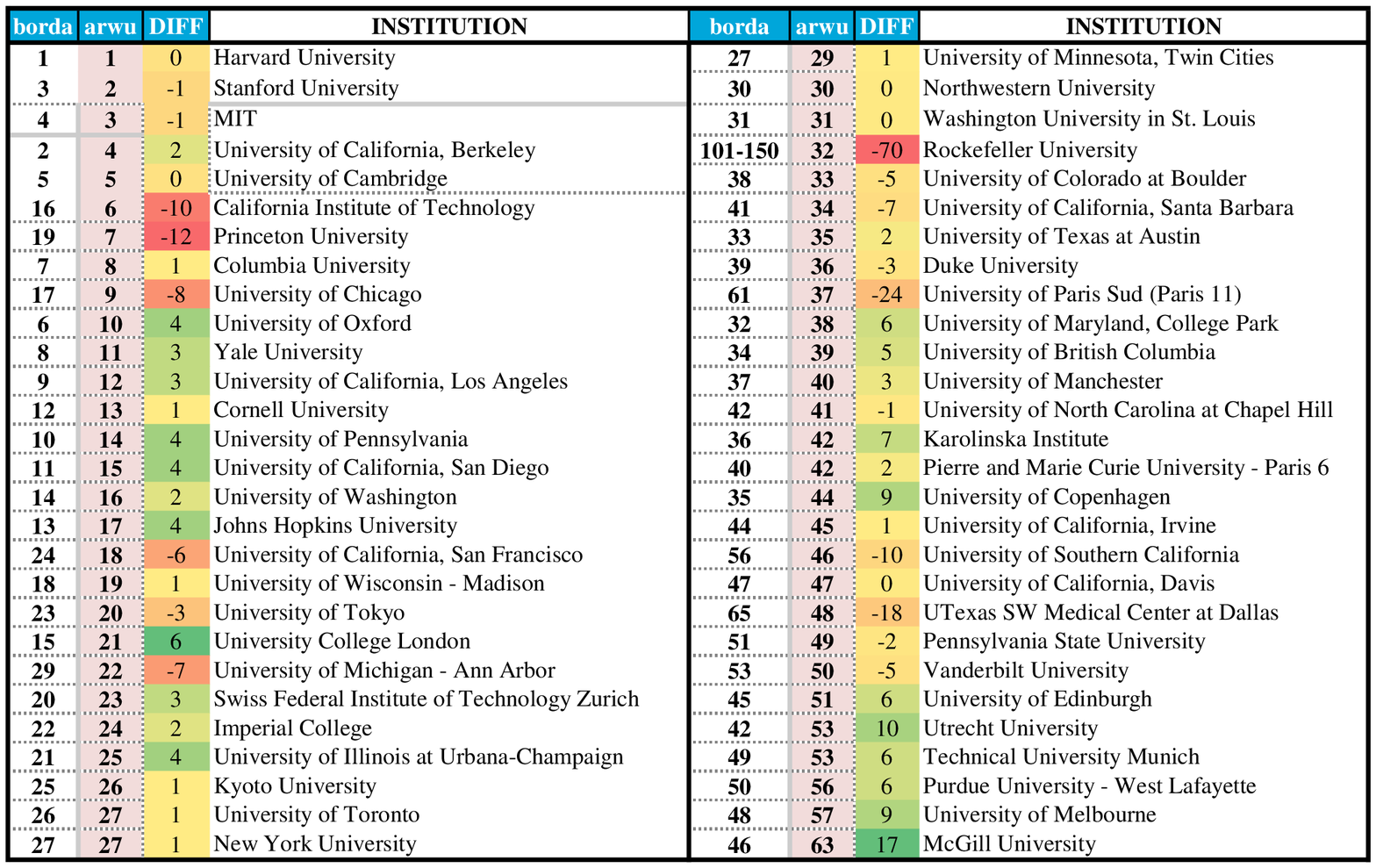}
\vspace{-3cm}
\caption{First tier  ranked using KAM in ARWU.\label{fig:50arwu}}
\end{figure}

\noindent Figure \ref{fig:50arwu} shows how the first tier of ARWU ranked universities would be shuffled were the KAM procedure used. According to the KAM procedure the $295$ institutions with a $0$ score for the indicator Alumni would be assigned a score of $5.90$; the $362$ institutions scoring $0$ on the Award indicator would be assigned a score of $7.24$; the $82$ institutions with no Highly Cited researchers among their faculty would be assigned a score of $1.64$, and the $10$ institutions with no papers in Science or Nature between 2007 and 2011 would be assigned a score of $0.2$ for this indicator.

\vspace{-3cm}

\begin{figure}[htbp]
\includegraphics[height=16cm,scale=0.5]{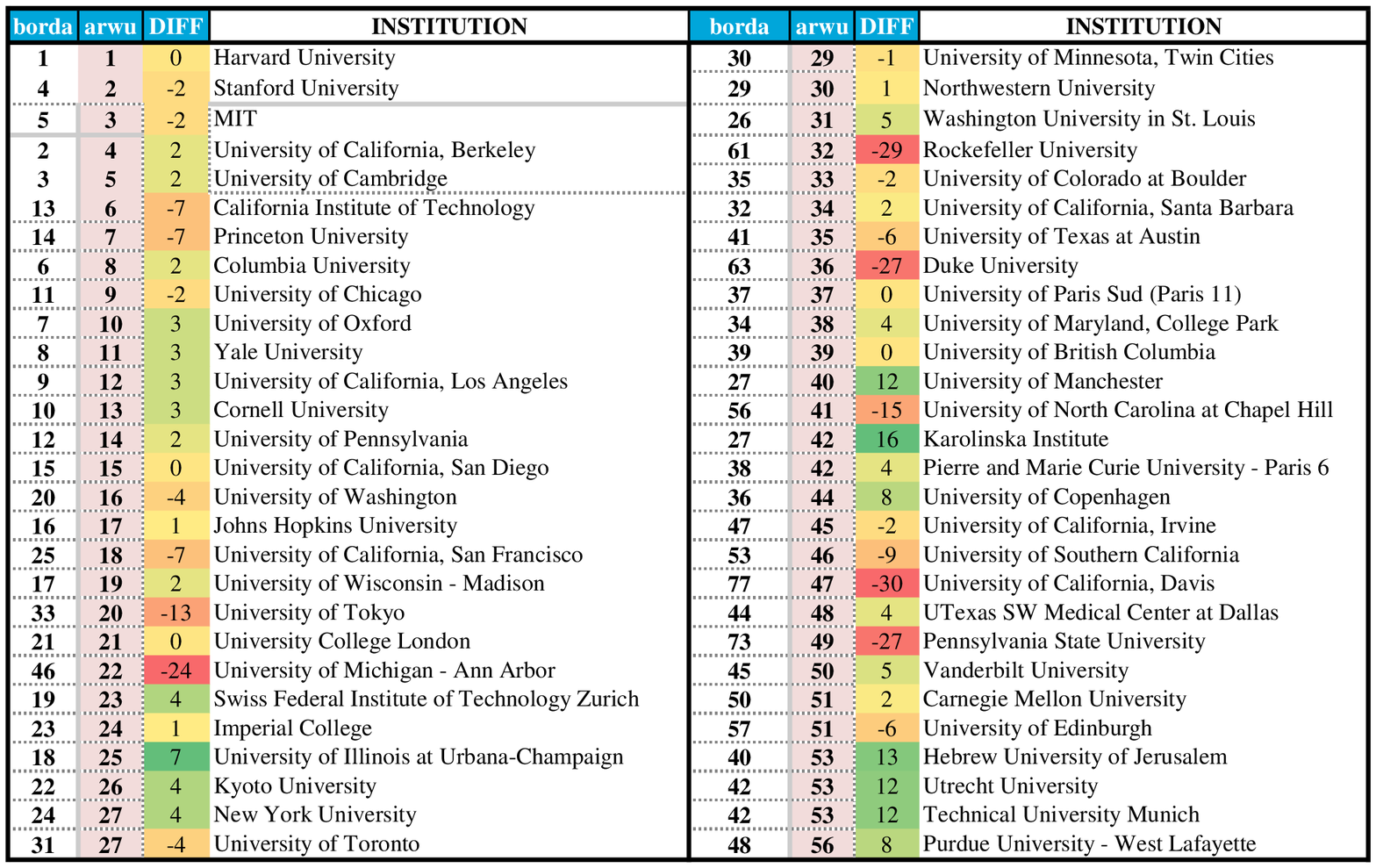}
\vspace{-3cm} 
\caption{\label{fig:remodeled} First tier  ranked using the remodeled KAM in ARWU.}
\end{figure}

\noindent The procedure leads to a relative advantage for those institutions with no previous ARWU score in an indicator. Small differences in the indicator PUB could also be amplified, a side effect of the dependence on irrelevant results. For institutions currently ranked outside the top $150$ the average of the changes in ranking (either up or down) would be $27$, with a maximum of $133$. These shifts could be reduced by rearranging the KAM rule in the following way:
 \begin{enumerate}
\item For each indicator $i$, count the number of institutions, $N$, with a score different from $0$.
\item Then for a particular university, identify the number of institutions with higher rank $(N_n)$.
\item Compute the score $I_n$ on the indicator following the normalization formula: $I_n = 10(1-N_n/N)$
\end{enumerate}
Although the (true) zeros have been restored and the perturbation introduced by the KAM versions of the Borda rule mitigated, the reshuffling would still be quite noticeable, as Figure \ref{fig:remodeled} shows. Institutions ranked outside the top $150$ exhibit an average change in ranking (either up or down) of $21$, with a maximum shift of $121$. In essence, the difference between ARWU rank and a rank generated by rank-based sequencing reflects the importance of the actual distribution functions of each ARWU indicator in the calculation of the total score.

\vspace{-3cm}

A KAM-like scoring system operates on a defined and complete population. However, other institutions might find a place in the ranking with either KAM-like scoring systems, adding more side effects and further reshuffling the results of the ranking.

\subsection{\label{combined} Score-driven sequences}

Score-driven sequencings aggregate indicator scores rather than the ranking of such scores. Scores of different events must be converted to a common currency or num\'{e}raire, and the rules for combining converted scores carefully constructed. In combined event athletics, for example, conversion attempts to ensure that similar performances in different events are rewarded alike, whereas in ARWU different indicators receive different weights in an attempt to reflect their respective importance in the final ranking. Dynamical operations such as gain, offset and power laws can yield modified indicators that are combined to yield an accepted score-driven ranking. Provided that the `exchange rate' among the different events is perceived as fair, particularly by the ranked subjects, the success of the ranking is essentially guaranteed.

A gold standard in the realm of score-driven sequencing of individual achievement is the ranking system used for combined events in track and field competition. Let us consider specifically the combined events scoring system adopted by the International Association of Athletics Federations in 1984 since it exemplifies most of the elements that may appear in score-driven tables \citep{IAAF}. Over more than a century of negotiations and investigations, the athletes themselves, officials, statisticians and sports scientists have arrived at a scheme for calculating a single combined score from the results obtained in a set of quite different events. The decathlon scoring system has been designed so that ``results in different disciplines with approximately the same point value should be comparable as far as the quality and difficulty of achieving those results are concerned" \citep{trkal}. To achieve that equalization (in much the same way as an equalizer works with recorded music) the scoring procedure for combined events uses two linear operators (offsets, $b_i$, and gains, $a_i$) and a non-linear element (powers, $c_i$), where the index $i$ refers to each of the $10$ events in the decathlon.

Let $m_i[n]$ and $s_i[n]$ represent respectively the mark (unmodified result in a discipline) and the score of athlete $n$ in event $i$. Figure \ref{fig:scheme} shows the transformations the mark $m_i[n]$ must go through to produce the score $s_i[n]$.

\setlength{\unitlength}{0.1cm}
\begin{center}
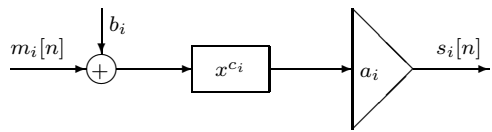
\begin{figure}[h]
 \begin{picture}(90,25)
  \put(30,10){\vector(1,0){10}}
 \put(30,12){$m_i[n]$}
  \put(42,10){\circle{4}}
    \put(40.65,9){$+$}
    \put(42,18){\vector(0,-1){6}}
     \put(43,15){$b_i$}
      \put(44,10){\vector(1,0){10}}
        \put(54,7){\framebox(10,6){$x^{c_i}$}}
              \put(64,10){\vector(1,0){11}}
  \put(75,2){\line(0,1){16}}
    \put(75,18){\line(1,-1){8}}
 \put(76,9){$a_i$}
    \put(75,2){\line(1,1){8}}
             \put(83,10){\vector(1,0){10}}
       \put(86,12){$s_i[n]$}
 \end{picture}
 \caption{\label{fig:scheme} Dynamics of the Decathlon scoring system}
 \end{figure}
 \end{center}

In the first summation of Figure \ref{fig:scheme}, the sign is positive for the mark $m_i[n]$ and negative for the offset $b_i$ in jumping and throwing events where a larger mark is superior. The signs are reversed in running events where a smaller mark is superior. After the offset is applied higher scores correspond to better performances. The offsets in the decathlon are also the threshold for scoring points i.e. the value that would receive $0$ points in each discipline. For instance, in the 100 m race the IAAF assigns $0$ points to any performance in excess of $b_i=18$ seconds and in the long jump any mark below $b_i=0.75$ m would receive $0$ points \citep{IAAF}. Although the offset introduces non-linearity by assigning zero points to all the results outside the threshold, there is no practical consequence since no athlete will perform outside the event threshold.

The power-law and gain elements in Figure 3 are tuned to try to guarantee similar rewards for performances of like quality across the full range of performance. However, the performance required in the three types of decathlon event (running, jumping and throwing) are not easily comparable, since athletes' achievements are affected by the conflict between anatomical features such as body weight and limb proportions, and physiological muscle-fibre-type requirements \citep{vandamme}. The fair assignment of scores to performances in such different events is thus far from straightforward. By calling on a huge amount of data collected at local, national and international track and field meetings, over many decades and with many different proposed tables, the IAAF has built a consensus around the importance of reflecting sensitively the {\em progressive} character of athletic effort. An improvement in the mark for an event becomes more difficult the closer the athlete comes to the maximum performance ever achieved and it is acknowledged that an acceptable scoring table should reward this effort with relatively larger increments at the highest levels of performance \citep{trkal}. This feature is known as {\it progressiveness} in the table.

However the inferred degree of progressiveness is not the same for the different type of athletic efforts required in decathlon disciplines. Consequently the IAAF has endorsed many adjustments leading to the current values of the gains and power laws. Today's values have been in place for almost 30 years with minor adjustments as measuring equipment and standards in track and field improve. They are not contested by athletes and federations, although the scoring procedure is not devoid of expert criticisms (\citeauthor{westera}, \citeyear{westera}; \citeauthor{geese}, \citeyear{geese}). Alternatives to the IAAF algorithms have also been explored, usually not to replace the tables but rather to exhibit physical drivers of scoring systems that would closely follow the IAAF tables. An example is the work of  \cite{grammaticos} based upon physical considerations combined with the use of Harder's methodology \citep{harder}.

As for the current system in place for grading decathletes, once the current values of the offsets $b_i$, gains $a_i$, and power indices $c_i$   are known for each discipline, it is straightforward to compute points $s_i$ corresponding to the mark $m_i$ expressed in seconds (running), meters (throwing) or centimeters (jumping) following Figure \ref{fig:scheme} \citep{IAAF}.
\begin{equation}
s_i=a_i{\mid m_i-b_i\mid}^{c_i}
\label{gains}
\end{equation}

\noindent All discipline scores $s_i$ are rounded to the nearest integer before being aggregated to produce the final event score of the decathlete,
$$S=\sum_{i=1}^{10}s_i$$

\noindent Unlike a rank-driven sequence, a score-driven sequence pays very close attention to the actual measurements of performance. Transformations of performance measurements within  a particular discipline are designed to produce a combined score that satisfies several objectives, notably that (i) comparable achievements in different disciplines should receive equal reward, and (ii) improvements in scores in progressive (regressive) events should require harder (easier) performance improvements as best-ever performance is approached.

\section{Should ARWU be score-driven or rank-driven?}

ARWU can be properly classified among the score-driven tables, but might it perhaps be preferable to employ a rank-driven procedure? The choice evidently turns on the meaningfulness of the actual indicator values as measurements of performance \citep{sawyer}, and on their suitability for aggregation. A score-driven procedure is indicated if the indicator measurements have meaning and can be combined, although the procedure is contingent of finding conversion strategies to equalize between indicators. If this is not the case, a rank-driven procedure should be preferred.

ARWU scores involve non-linear transformations and normalization by the best performer \citep{docampo2013}. However, to discuss the meaningfulness and suitability of the data for aggregation purposes we start from the raw data before any linear or non-linear operation takes place. To explore the raw data for the possible presence of underlying phenomena we apply principle components analysis (PCA) along the lines of \cite{dehon} and \cite{docampo2011}. The raw scores on the five non-composed indicators present a high degree of correlation, as Table \ref{tablauno} shows for the 2012 ARWU edition. The PCP indicator is excluded because it aggregates the other five indicators.

 \begin{table}[htbp]
  \centering
     \begin{tabular}{lrrrrrr}
    \hline
    \multicolumn{7}{c}{\textbf{Correlation Matrix}} \\
    \hline
    \multicolumn{2}{c}{ } & \multicolumn{1}{c}{ALUMNI} & \multicolumn{1}{c}{AWARD} & \multicolumn{1}{c}{HICI} & \multicolumn{1}{c}{S\&N} & \multicolumn{1}{c}{PUB} \\
          & \multicolumn{1}{r}{ALUMNI} & 1.0 & .873  & .733  & .797 & .526 \\
          & \multicolumn{1}{r}{AWARD} & .873  & 1.0 & .765  & .790 & .458 \\
    Correlation & \multicolumn{1}{r}{HICI} & .733  & .765  & 1.0 & .929  & .699 \\
          & \multicolumn{1}{r}{S\&N} & .797  & .790  & .929  & 1.0 & .701 \\
          & \multicolumn{1}{r}{PUB} & .526  & .458  & .699  & .701 & 1.0 \\
            \hline
    \end{tabular}%
  \caption{Correlation Matrix for the non composed raw indicators of ARWU}
  \label{tablauno}%
\end{table}%

\noindent The size of our sample is appropriate for Principal Component Analysis:  \cite{tabachnick} suggests that a number of samples in excess of $300$ is comfortable for Factor Analysis. Inspection of the correlation structure of the five non composed indicators from Table \ref{tablauno} reveal that all the correlation coefficients are highly significant (at the $0.001$ level, 1-tailed). The Kaiser-Meyer-Olkin value of the 2012 ARWU sample is $0.8$, far exceeding the recommended value of $0.6$ \citep{kaiser}. Bartlett's Test of Sphericity \citep{bartlett} reaches statistical significance. Both results stand in support of the factorability of the correlation matrix.

To examine the properties of a scale based upon the first five ARWU indicators, the raw data from the $500$ world-class universities were examined for Principal Component Analysis using SPSS.  Since the raw measures are not commensurable, performing the component analysis on the correlation matrix is preferable for statistical reasons (\citeauthor{Morrison}, \citeyear{Morrison}; \citeauthor{stevens}, \citeyear{stevens}). It is worth recalling that PCA is just a technique to reduce the dimensionality of a data set in search of the underlying constructs that account for most of the variance of the data set --- a percentage in excess of 75\% \citep[page 327]{stevens}.  The analysis reveals just one component with an eigenvalue exceeding $1$, explaining a great deal of the variance in the sample, as shown in Table \ref{eigen}.

\vspace{-0.25cm}
 \begin{table}[htbp]
  \centering
            \begin{tabular}{rrrr}
    \hline
    \multicolumn{4}{c}{\textbf{Total Variance Explained}} \\
    \hline
    Eigenvalue & \multicolumn{1}{c}{value} & \multicolumn{1}{c}{\% of Variance} & \multicolumn{1}{c}{Cumulative \%} \\
    1     & 3.93 & 78.65 & 78.65 \\
    2     & .64  & 12.75 & 91.4 \\
    3     & .25  & 4.99 & 96.4 \\
    4     & .12  & 2.34 & 98.77 \\
    5     & .06  & 1.23 & 100.0 \\
    \hline
    \end{tabular}%
    \caption{\label{eigen} Percentage of variance explained by the five components}
\end{table}

As expected, given the values shown in Table \ref{eigen} the scree test reveals just one significant eigenvalue. As noted by \cite{hakstian}, for a sample of more than $250$ items and a mean communality in excess of $0.6$, as it is the case of the ARWU sample, the scree test yields an accurate estimator of the number of factors, one in this case, since the ratio of the number of factors, $1$, and the number of indicators, $5$, is $<0.30$. The communalities explained by this first factor (Table \ref{loading1}) imply that great deal of the variance in the five indicators is adequately accounted for by the first principal component. Four of the five values of the communalities are far greater than $0.6$, the fifth one not far from the mark, guaranteeing the reliability of
the first principal component \citep{guadagnoli}. According to \cite{comray},  loadings in excess of $0.71$ are considered excellent,  as it is the case of the five loadings shown in Table \ref{loading1}.

 \begin{table}[htbp]
  \centering
        \begin{tabular}{||c|c||c|c||}
            \hline
\multicolumn{4}{||c||}{First Principal Component }\\\hline
\multicolumn{2}{||c|}{Loadings} & \multicolumn{2}{|c||}{Communalities}                                 \\
\hline
Indicator	&	Loading		&	Initial	&	Extraction		\\ \hline\hline
AlUMNI	&	0.892	& 1.0	&	0.796	\\ \hline
AWARD	&	0.885	& 1.0	&	0.783	\\ \hline
HICI	&	0.936 	&  1.0	&	0.876	\\ \hline
N\&S	&	0.957	&	1.0	&	0.915			\\ \hline
PUB	&	0.750	&	1.0	&	0.563	\\
\hline
     \end{tabular}
            \caption{\label{loading1} Loadings of the five indicators on the first Principal Component}
\end{table}

The loadings on the first component exhibited in Table \ref{loading1} reveal the degree of correlation between each indicator and that component. They are significant for all indicators. Moreover, the approximate equality of the loadings suggests that a balanced weighting system could be used to combine the normalized raw data (adjusting for the different means and standard deviations within each indicator).

The analysis of the factor loadings can be carried  further by considering the standard errors of the estimates, although  factor loading error estimation error is under-researched \citep{sass}. While it would be possible  to use re-sampling techniques to bootstrap the analysis \citep{zientek},  we have just one component and there are therefore no complex cross effects between orthogonal components. \cite{cliff} suggests that the standard error in the factor can be estimated by doubling the standard error corresponding to ordinary correlation. However, given the large amount of ARWU data and the small number of indicators and components, this is clearly too conservative. Indeed, when the sample size is large relative to the number of variables the factor standard errors are generally no larger than $1.5$ times the correlation error \citep[page 332]{stevens}. For the ARWU sample the standard error corresponding to an ordinary correlation would be $\sigma=1/\sqrt{499}=0.045$. Table \ref{nominal} shows the nominal values of score coefficients obtained from the factor loadings, along with $95\%$ confidence intervals for the estimates based upon a standard error $1.5$ times  the canonical $\sigma=0.045$:

\vspace{-0.25cm}
  \begin{table}[htbp]
  \centering
        \begin{tabular}{||c|c||c|c||}
            \hline
\multicolumn{4}{||c||}{First Principal Component }\\\hline
\multicolumn{2}{||c|}{Score Coefficients} & \multicolumn{2}{|c||}{$95\%$ confidence intervals}                                 \\
\hline
Indicator	&	Score Coefficient		&	lower bound	&	upper bound		\\ \hline\hline
AlUMNI	&	0.23	& 0.19	&	0.26	\\ \hline
AWARD	&	0.23	& 0.19	&	0.26	\\ \hline
HICI	&	0.24 	&  0.20	&	0.27	\\ \hline
N\&S	&	0.24	&	0.21	&	0.28			\\ \hline
PUB	&	0.19&	0.16	&	0.22	\\
\hline
     \end{tabular}
            \caption{\label{nominal} Loadings of the five indicators on the first Principal Component}
\end{table}

As the results in Table \ref{nominal} show, equal score coefficients would lie within the five confidence intervals. In practice, however, ARWU halves the relative weight applied to the Alumni indicator, and assigns the other half to the size compensator (indicator PCP). This modification might reflect a view of the ARWU authors that the construct validity of the indicator Alumni with respect to the quality of education is low, as pointed out by \cite{ioannidis}. The authors may also feel that a combined weight of $0.4$ assigned to the two indicators related to Nobel Prizes and Field's Medals would be too high.

We conclude that the raw ARWU data conforms to an underlying one-dimensional scale, that the indicators are well suited for weighted aggregation, and that the weights adopted by ARWU are acceptable in statistical grounds. While a rank-driven procedure could be used to aggregate ARWU indicators, information would be lost and it is preferable to use the adopted score-driven procedure.

\section{Analysis of ARWU as a score-driven table}

As we have seen, the consistency of the Shanghai ranking indicators suggests that a score-driven procedure is appropriate, contingent on finding conversion strategies to harmonize the scores. We know that the authors of ARWU have adopted a uniform, non-linear compression (square root law) for all of the indicators: see \cite{docampo2013} for a complete explanation of the internal operations of ARWU. The gains used to equalize the indicators are best-performer dependent, so in every edition of ARWU the gains change as a result of the chosen methodology. However, given the stability of the raw scores of the best performers in the previous editions of ARWU, the change is relatively small from year to year.

A comparison with the combined events scoring system analysed in Section \ref{combined} will provide considerable insight into the dynamics of the Shanghai ranking. In practice the Shanghai ranking procedures follow similar guidelines to the IAAF scoring system, although obviously with a different set of parameters and non-linear conversions. Let $m_i[n]$ and $s_i[n]$ be the raw points and the ARWU score on the indicator $i$ of university $n$, respectively. Certainly no inversion of the type used in the running events in track and field is required since all the scores in ARWU point in the same direction. No other offsets are needed for the first four indicators, since the minimum  raw score is $0$. Perhaps an offset might be considered for the PUB indicator, but  this would open questions relating to institutional size. Therefore, there is no need in ARWU for the first addition stage in Figure \ref{fig:scheme}.

Consequently the offsets are set to $0$ and the powers to $0.5$  for all the indicators. For the computation of the gains it is necessary to first identify the best performer on each indicator. Let then $BP_i$ be the raw score achieved by the best performer on the indicator $i$. The gain would then be   \begin{eqnarray}
   a_i& = & \frac{100}{\sqrt{BP_i}},\;1\le i\le6.\label{gaina}
   \end{eqnarray}

The gains in ARWU have been chosen to equalize the peak performance on each indicator. Had the gains been chosen instead to equalize the average scores on each indicator, then obviously the gains of the indicators in which the scores are non-zero for most universities (PUB and to a lesser extent S\&N) would have taken a lower value. This would be inconvenient, given the gap already created by the first two indicators. The price ARWU pays for this simplification is that the scores become best-performer dependent. We show in the next section that ARWU could expunge this undesirable feature.

We conclude this section by stating that ARWU is a truly score-driven table with many features also found in  the IAAF combined events scoring system. The difference, besides the values of the parameters, is that the gains in ARWU change on a yearly basis, according with the performance of the highest ranked university on each indicator.

\section{Removing the best performer-dependence in ARWU}

The normalisation of each ARWU indicator serves only to tie down the meaning of the relative indicator weights. Without some form of normalisation the meaning of the weights would evolve capriciously. However the normalisation does not need to be to the peak in any year, nor does it need to vary from year to year to achieve this aim. Indeed in the decathlon the increase of points corresponding to the world record (from around 8,000 in 1955 to more than 9,000 in 2012) is an important feature of the cultural history of the sport, and one that could be usefully reflected in the ARWU procedures.

By adopting a normalisation to the peak for the year in question great importance is attached to the annual variation of performance of Harvard University which sets the yearly normalization gains in 5 out of 6 indicators. Therefore, a lower (higher) performance of Harvard in a particular indicator means that the ARWU ranks are rebalanced to apply a higher (lower) effective weight to the population on that indicator. This dependence could be alleviated if a number of universities were used to establish the picture of relative shifts from one indicator to another. As a matter of fact this possibility should be explored by ARWU to cope with the recently announced changes in the methodology for identifying highly cited researchers \citep{thomson2012a}.

One way of avoiding the best-performer dependence would be to tie down the normalisation factors by considering a basket of universities and averaging over a few years, and  to use those factors to normalise every year in the future. This would provide the means for year-on-year comparison of global performance improvement by the world's universities, something of considerable interest to stakeholders. It would be like tracking the inexorable improvement of human sporting achievement by competing against a world record that can be repeatedly broken.

To show that a procedure with fixed gains is workable with the minimum reshuffling of the ARWU list, let us first identify the first five indicators:
\begin{center}
\begin{verbatim}
ALUMNI=1; AWARD=2; HICI=3; S&N=4; PUB=5
\end{verbatim}
\end{center}
Again, let $m_i[n]$ be the raw score for university $n$ in the indicator $i$. For example, a university with 2 affiliated HCR will have $m_3[n]=2$ and a university with one paper having a first author on Science or Nature during the period of analysis will have $m_4[n]=0.5$.

We will set the gains $a_i$ for the first five indicators by averaging the actual gains used in 2011 and 2012:
 $$a=(a_1,\ldots,a_5)=(17.875,\;	16.975,\;	7.225,\;	4.775,\;	0.850)$$
 Now, following the scheme described in Figure \ref{fig:scheme}, we rewrite Equation \ref{gains} as follows:
 \begin{equation}
s_i[n]=a_i{m_i[n]}^{\frac{1}{2}}
\label{scores}
\end{equation}

\noindent The computation of the PCP indicator following the general procedure outlined in Figure \ref{fig:scheme} necessitates distinguishing between cases where FTE is known or unknown. Let us begin with the majority of universities (more that 80\% of the institutions included in the ARWU list) for which FTE data are used in ARWU. For those universities, let  $FTE[n]$ be the number of full-time equivalent faculty of university $n$. We need to generate the `signal' $m_6[n]$ to activate the operations in Figure \ref{fig:scheme}. \cite{docampo2013} shows that raw scores, $m_6[n]$, on the PCP are obtained as follows:
\begin{equation}m_6[n]=\frac{1}{FTE[n]}\left(a_1^2m_1[n]+2\displaystyle{\sum_{i=2}^{5}a_i^2m_i[n]}\right)\label{pcpavge}\end{equation}
In the case of universities focused on the social sciences, the weights are a little different:
$$m_6[n]=\frac{1}{7FTE[n]}\left(9a_1^2m_1[n]+14\displaystyle{\sum_{i=2}^{5}a_i^2m_i[n]}\right)$$
From there, we use the average gain used in 2011 and 2012, $a_6=9.325$, and apply Equation \ref{scores}.

When the data on FTE are not available for an institution, it can be shown that ARWU procedures are equivalent to assigning a single dummy FTE number to all such institutions. The computation of the PCP indicator in ARWU is based upon the scores on the five indicators according to the procedure described in  \citep{docampo2013}. We summarize here the details: First, the weighted squares of the scores are added:
\begin{equation} {\rm WS}=0.1{\rm ALUMNI}^2+0.2({\rm AWARD}^2+{\rm HICI}^2+{\rm N\&S}^2+{\rm PUB}^2)\label{wss}\end{equation}
Let $CAL$ be the quotient of $WS$ and the $FTE$ of the best performer (Caltech in 2011 and 2012). Then:
\begin{enumerate}
\item If the FTE number is known, then: ${\rm PCP}=100\sqrt{\frac{WS/FTE}{CAL}}$
\item If the FTE number is not known, then: ${\rm PCP}=\sqrt{\frac{WS}{K}}$
 \end{enumerate}
The parameter $K$ is not discussed in the ranking explanatory documentation, although its values has been found empirically from a regression analysis on the actual PCP scores in ARWU 2011 \citep{docampo2013}. In 2011, $K=0.94$. We have done the same regression analysis using the data from 2012 and found that the value of $K$ changed to $1.075$. Over the years of existence of ARWU, the value of this parameter has been consistently set around $1.0$, but has never remained constant.

Now, the dummy value of FTE can be recovered from the two formulas for PCP (with and without explicit FTE numbers), as follows:
\begin{eqnarray}
{\rm PCP}& = & 100\sqrt{\frac{WS/FTE}{CAL}}\nonumber\\
& = & \sqrt{\frac{WS}{K}}\Rightarrow\nonumber\\
FTE & = & 10^4\frac{K}{CAL}\label{fte}
\end{eqnarray}
Now, by using the values of $K$ and $CAL$ in 2011 and 2012 we get the dummy values of FTE, $822$ and $955$, respectively. We thus adopt in Equation \ref{pcpavge} the value $890$ as the FTE number of universities for which explicit data on FTE are not available.

Finally, we use the weights adopted in ARWU to aggregate the final score, $s[n]$, for standard institutions or the ones concentrated on the social sciences, respectively:
\begin{eqnarray*}
s[n]& = &\frac{1}{10}\left(s_1[n]+s_6[n]+2\sum_{i=2}^{5}s_i[n]\right)\\
s[n]& = &\frac{1}{10}\left(1.25s_1[n]+1.25s_6[n]+2.5\sum_{i=2}^{5}s_i[n]\right)
\end{eqnarray*}

\begin{figure}[t!]
\includegraphics[height=22cm,scale=1]{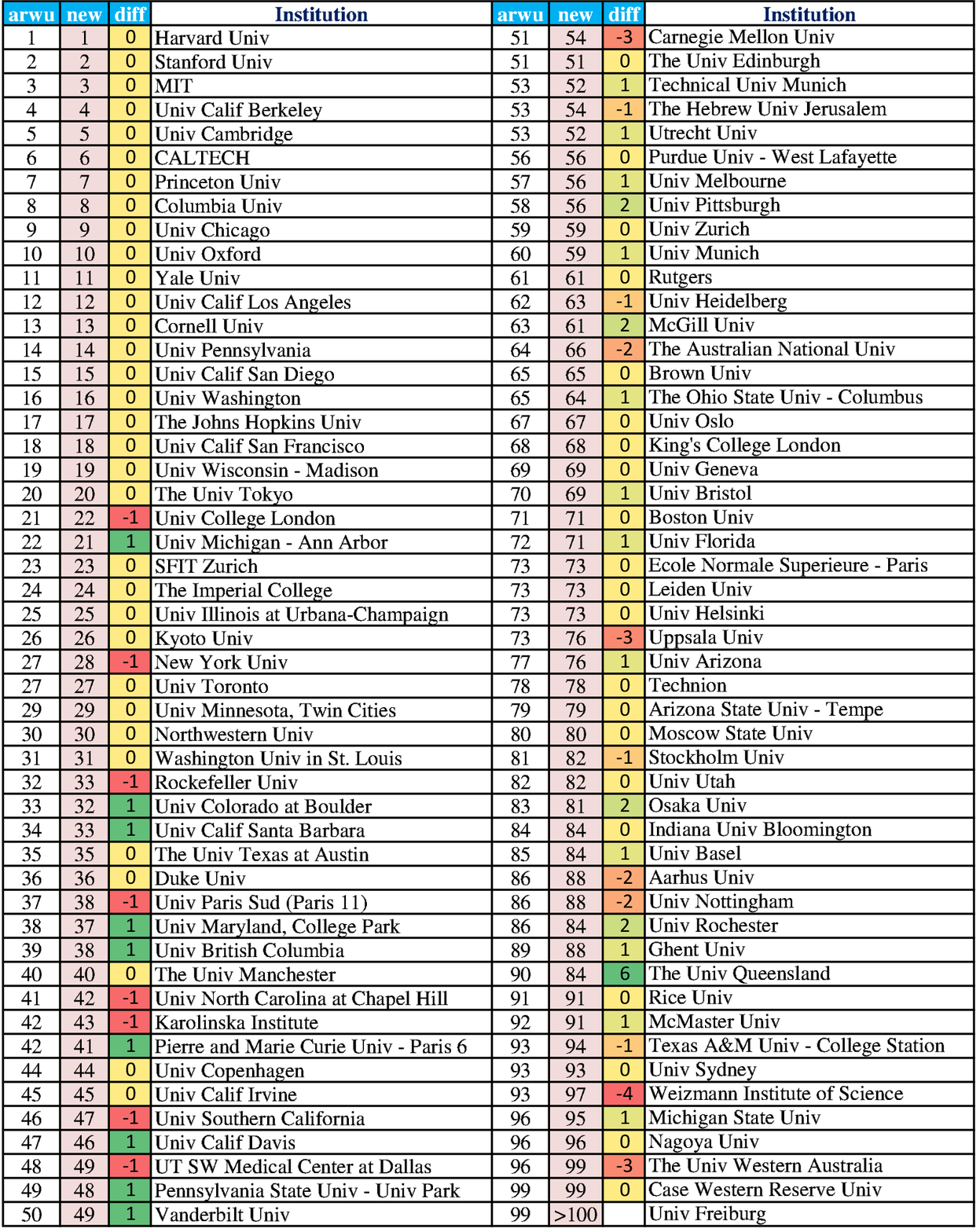}
\vspace{-7cm}\caption{\label{fig:oneh} First tier of ARWU with the actual ranks (arwu) and the ranks when using the suggested fixed gains (new).}
\end{figure}

If that procedure had been used in ARWU 2012, the results for the first 100 universities would have almost been the same, since the gains chosen are close to the ones actually used in 2012 (see  Figure \ref{fig:oneh}). The suggested scoring procedure would  reveal the year-on-year evolution of the performance of all the universities in the ARWU list. Given the stability of the best performers in the past, no further adjustments of gains would be necessary. If the gains were held at agreed constant values, any university could in principle calculate its ARWU indicator scores (but obviously not rank) without recourse to the delays and complexities introduced by ARWU processing.

\

\section{Discussion}
We have shown how the Shanghai ranking can be correctly classified as a score-driven table. We have also suggested that ARWU involves regressive indicators, and now turn to a deeper analysis of this question. Specifically, we conjecture that every non-composed sub-scale of ARWU constitutes a regressive indicator and that this is appropriate, as distinct from IAAF combined events tables that have constructed to be mildly progressive.

Physical human performance, so contingent on anatomical and physiological constraints, is arguably characterized by progressively greater difficulty in improvement close to the highest achieved performance limits. On the other hand, research performance  is arguably characterized by regressively lower difficulty in pushing the limits.  In each of the ARWU non-composed indicators, a substantial number of institutions at the top achieve high raw scores that are nevertheless separated by significant margins of difference. Below the top echelon, there is a larger number of very good achievements separated by narrower margins of difference.  Just as physical constraints shape the progressive character of human achievement, it is likely that resource constraints shape the regressiveness of university achievments. At the institutional level, outstanding research achievements justify and attract additional investments in research infrastructure and foster team-building and collaboration within the institution, driving ever higher performance. Moreover, institutional performance is composed of the efforts of individuals and teams, and abundant resources act as a magnet for outstanding researchers further intensifying institutional performance.

Ultimately, the regressive or progressive character of the scoring system for any competition aims to signal the degree of difficulty faced by a contestant who strives to improve in points and rank. While a small improvement in a raw score in a progressive sporting event does not  lead to substantial rank advancement except the top of the ranking, publishing even one or two more papers in Science or Nature may represent a significant advance for a low-ranked university. Conversely, a high-ranked university will not advance far with one or two more papers: as a matter of fact, it may not advance at all.

To understand further the regressive character of the ARWU indicators, let us formalize some concepts. In the first step it is important to attend to the raw indicator values, since the dynamical conversion of these to the final score is precisely the point of our study. For indicator $i$ and university $n$ define the raw score $s(i,n)$. For any indicator $i$, obtain the rank statistics: i.e. sort by raw scores so that $s(i,n) \ge s(i,n+1) \forall  n$.

\begin{center}
\begin{figure}[t]
\includegraphics[totalheight=18cm,width=14cm,scale=2]{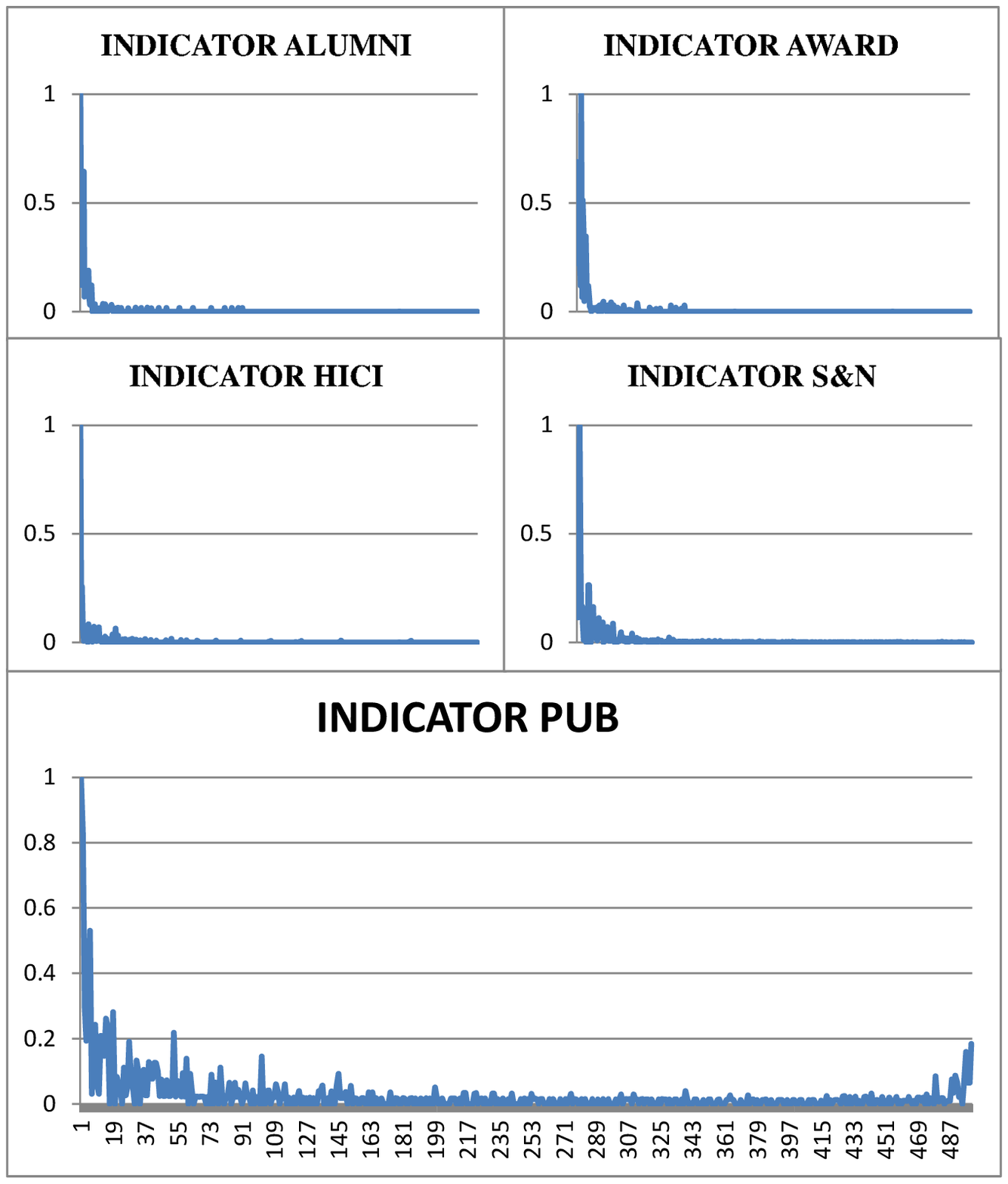}
\vspace{-7cm}
\caption{\label{fig:differences} Scoring difference function, $ds(i,n)$, for the first five indicators in ARWU.}
\end{figure}
\end{center}

Define $ds(i,n)$ as the difference in raw scores of adjacent ranked institutions, $ds(i,n)=s(i,n)-s(i,n+1)$. A regressive indicator will satisfy $ds(i,n)>ds(i,m), n < m$: the raw score increment required to attain the next highest rank is larger for better ranks. For our purposes some infrequent departures from this inequality would be immaterial, so we denote a regressive indicator as one with the inequality $ds(i,n)\ge ds(i,m),\; n < m$ holding for most $(n,m)$. For a regressive indicator, the function $f(n)=ds(i,n)$ should converge to zero as $n$ increases, albeit with some noise.

To examine $ds(i,n)$ for the ARWU indicators, we convert the five 2012 ARWU indicators to raw scores, using estimations of the raw scores of Harvard University and the findings in \citep{docampo2013}. The file with the raw scores of all the 500 universities is available upon request. Figure \ref{fig:differences} shows the function $f(n)=ds(i,n)$ for the first four indicators in ARWU. Harvard University is omitted because the value of $f(n)$ is so large that it masks the rest of the data. The plots have been re-scaled to equal maxima to remove any influence of small errors in the estimations of the scores of Harvard University.
As Figure \ref{fig:differences} shows, the first four ARWU indicators are regressive and the envelope converges uniformly to 0 as the rank order position $n$ increases. 

However, the PUB indicator is anomalous since a cluster of the lowest-ranked institutions do not converge towards $0$. This phenomenon reflects the fact that institutions with unremarkable but diverse raw PUB scores but  non-zero scores in any of the other four indicators may find a place in ARWU. We believe that the effect results from the way ARWU selects universities to enter the ranking, and is not a property of the PUB indicator itself. Had we used just the first 400 world class universities the picture for PUB would have looked like the others.

\begin{center}
\begin{figure}[h!]
\includegraphics[totalheight=16cm,width=12cm,scale=1]{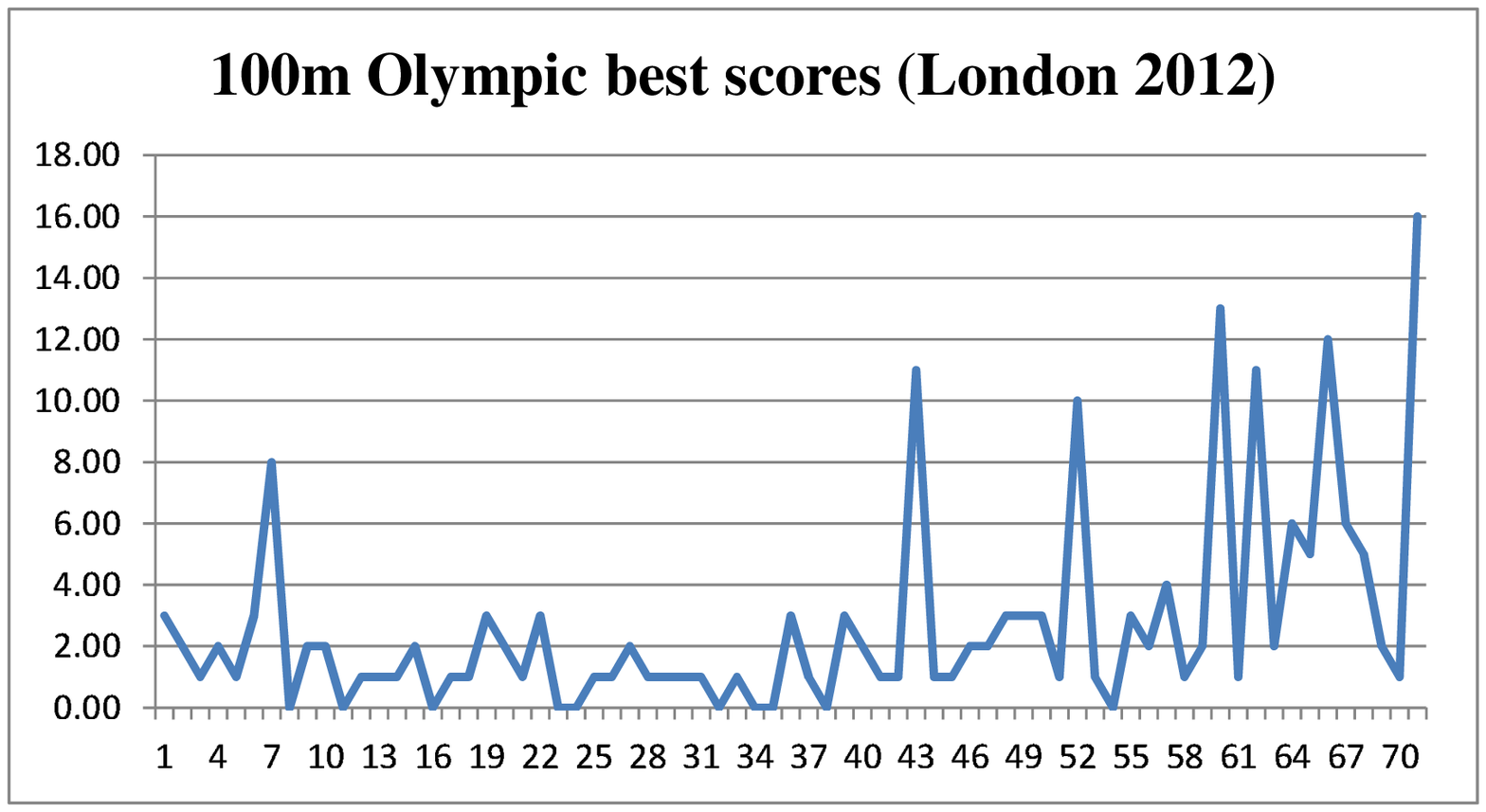}
\vspace{-10cm}\caption{\label{fig:olympics} Scoring difference function, $ds(n)$, for the best qualifying scores of the participants in the 100m at the London Olympics (2012). The y-axis unit is one hundredth of a second. }
\end{figure}
\end{center}

It is of interest to relate the regressive character of the information provided by Figure \ref{fig:differences} to a progressive measure. Figure \ref{fig:olympics} exhibits the difference function $ds(n)$ for the best scores obtained by the participants in all the qualifying 100 meters  2012 London Olympics races \citep{olympics} leading to the Olympic final held on August 5th, 2012.  As expected for an event involving peak human performance,  even substantial margins of improvement at the bottom of the rank do not purchase significant advances in the qualifying ladder.

Having established that the raw ARWU indicators are regressive, we close the discussion with the conjecture that the ARWU square-root compression operation (`statistical adjustment' in the words of the ARWU rankers) has been adopted as a dynamical transformation designed to {\em de-emphasize} the regressive character of the raw indicators. Perhaps the motivation is to encourage institutions with modest raw scores, and to enhance their buy-in to a ranking system that they might otherwise resist should it appear to be a bridge too far.
\section{Conclusions}

We have established that ARWU ranking is properly a score-driven ranking system, and we have identified the meaning and numerical values  of the gains and power laws expected in such a scoring system. We have also examined the ARWU indicator scores as sub-scales of a measure of research performance and found them to be so coherent that the underlying aggregate scale is one-dimensional. Using Principal Component Analysis, we have shown that the first principal component explains a substantial proportion (in excess of 78\%) of the variance in the raw scores of the sample of five hundred world class universities.

We have explained the role of the non-linearity used in the Shanghai ranking (the compression of the scores before normalization) and have also explored its suitability for the ranking data. In doing this, we have developed a mathematical formulation to distinguish regressive from progressive indicators. Based upon that formulation, we have shown that the non-composed indicators of ARWU constitute regressive measures of performance, as opposed to the indicators in the combined events in track and field that have been devised to be of progressive character.

There is no agreed or objective measure of the quality of research performance at institutional level. If a measure is required, our findings  support the use of the ARWU indicators  as one set of measures of this quality, and of the ARWU ranking as a summary thereof. As discussed by \cite{sawyer} the process of measurement in the social sciences, when construct validity cannot be assessed by benchmarking, necessarily includes social processes of persuasion and convergence alongside operational processes of estimation. An objective of social science measurement is to ensure that others acknowledge the measures as benchmarks replacing the unknown and unknowable measure. If the measure is widely accepted, then `individuals and entities will minimise risk by using the measure, rather than constructing a new measure' \citep{sawyer}.

In summary the internal dynamics of the ARWU ranking correspond to a well designed score-driven table in which the non-linear elements have been chosen appropriately in light of the regressive nature of measures of research performance. Operationally any measurement error associated with the six indicators that compose ARWU can be controlled.  Indeed, readers may experiment with the implication of measurement uncertainties by applying the methodology presented in section \ref{methodology}.

The direction of our studies will now turn to a deeper exploration of ARWU success factors by analysing how various university stakeholders use the ARWU ranking to advance their various interests. We hope to understand better the anchoring capacity of the ARWU ranking among the different measures that attempt to capture the research quality of a Higher Education institution.

\noindent\bigskip{\bfseries Methods Summary}

 \

 \noindent ARWU data on academic institutions were gathered directly from the Shanghai Jiao Tong University ARWU website, http://www.shanghairanking.com. Estimation of raw scores were computed using an Excel file containing the ARWU data and the raw estimates for Harvard University. The Principal Components Analysis was performed using SPSS Statistics $19.0$. All Excel and SPSS files are available upon request.


\end{document}